\documentclass{article}
\usepackage{amssymb}
\usepackage{bm}
\usepackage[dvips]{graphicx}
\begin{document}

\title{Switching magnetic junction by joint action of current pulse and magnetic field: Numerical simulation}

\author{S. G. Chigarev, E. M. Epshtein\thanks{E-mail: eme253@ms.ire.rssi.ru}, Yu. V. Gulyaev, P. E.
Zilberman\\ \\
V.A. Kotelnikov Institute of Radio Engineering and Electronics\\
    of the Russian Academy of Sciences, 141190 Fryazino, Russia}

\date{}

\maketitle

\abstract{The results are presented of a numerical simulation of the
switching magnetic junction by a spin-polarized current pulse under
applied magnetic field with the current density and field below the
threshold values. A possibility is shown of the switching with
controllable delay relative to the current pulse.}\\ \\

The switching magnetic junction by spin-polarized current~\cite{Katine} attracts
continuous attention both of experimentalists and theorists. This is due
to a possible practical application of the effect for high-density
recording information by current in magnetic media, as well as interesting
problems related with the attempts of theoretical interpretation of the
experimental results and prediction new effects.

There are two important problems in experimental investigation and
application of the effect, namely, lowering the switching threshold and
shortening the current pulses causing the switching. Solution of the
problems will allow decreasing the Joule heating of the junction layers
and avoiding destruction of the switching elements by the flowing current.

Nowadays several ways have been proposed to lower the threshold current
density, under which the initial configuration becomes unstable and
switches to another one with different electric resistance. These are
proper choice of the barrier layer material~\cite{Diao}, using diluted magnetic
semiconductors as the pinned and free layers of the magnetic junctions~\cite{Watanabe},
using combination of the layer parameters corresponding to efficient
spin injection from the pinned layer to the free one and ``locking up''
the injection at the interface between the free layer and the nonmagnetic
layer closing the electric circuit~\cite{Gulyaev1}, using a noncollinear initial
configuration of the magnetic junction~\cite{Brataas,Chung,Mankoff,Chigarev}.

In Ref.~\cite{Gulyaev2} attention was paid to a possibility of substantial lowering of
the current density needed to switch a magnetic junction under joint
action of the current and magnetic field with the current density and
field below the corresponding threshold values when the factors mentioned
act singly. If the magnetic field strength is close enough to the
threshold value (the anisotropy field), but slightly smaller than it, then
a current with density below substantially than the
threshold value without magnetic field can help to switching. Note, that the presence of the
magnetic field does not break local character of the exchange switching
the magnetic junction, because the magnetic field of such a value cannot
do switching alone (without current).

Together with the lowering current density of the stationary current, the
presence of the magnetic field opens possibility of switching by short
current pulses. Varying the pulse amplitude and duration allows switching
with a time delay relative to the current pulse.

Theoretical investigation of the switching by short current pulses is
related with solving a set of nonlinear differential equations (the
Landau--Lifshitz equations) with variable coefficients. Because of
difficulty of this task, it is need to use numerical solution.

Let us consider a magnetic junction with the free layer thickness small
compared to the spin diffusion length, so that the macrospin
approximation~\cite{Gulyaev2} is applicable. The high spin injection conditions~\cite{Gulyaev1} are assumed to
be valid, so that the switching magnetic junction is determined by
creating nonequilibrium spin polarization with injected spins~\cite{Heide,Gulyaev3}.
Under these approximations, the free layer magnetization in presence of
the time-depended current $j(t)$ perpendicular to the layers is described
by the following equation~\cite{Gulyaev2}:

\begin{eqnarray}\label{1}
  &&\frac{d\hat\mathbf M}{dt}-\kappa\left[\hat\mathbf M,\,\frac{d\hat\mathbf
  M}{dt}\right]+\gamma\left[\hat\mathbf M,\,\mathbf H\right]+\gamma
  H_a\left(\hat\mathbf M\mathbf n\right)\left[\hat\mathbf M,\,\mathbf
  n\right]\nonumber\\
  &&+\gamma\left[\hat\mathbf M,\,\mathbf H_d\right]
  +\gamma H_a\frac{j(t)}{j_0}\left[\hat\mathbf M,\,\hat\mathbf M_1\right]=0.
\end{eqnarray}

Here $\hat\mathbf M=\mathbf M/|M|$ is the unit vector along the free layer
magnetization vector $\mathbf M$, $\mathbf H$ is the external magnetic
field, $\mathbf H_d$ is the demagnetization field, $H_a$ is the anisotropy
field, $\mathbf n$ is the unit vector along the anisotropy field,
$\hat\mathbf M_1$ is the unit vector along the pinned layer magnetization,
$\kappa$ is the Gilbert damping constant, $\gamma$ is the gyromagnetic
ratio;

\begin{equation}\label{2}
  j_0=\frac{eH_aL}{\mu_B\alpha\tau Q_1}
\end{equation}
has the meaning of the threshold current density for the switching
magnetic junction by a stationary current (an infinitely long pulse) without
magnetic field; $\mu_B$ is the Bohr magneton, $\alpha$ is the \emph{sd}
exchange interaction constant, $\tau$ is the electron spin relaxation
time, $L$ is the free layer thickness, $Q_1$ is the spin polarization of
the pinned (injecting) layer conduction.

The last term in the left-hand side of Eq.~(\ref{1}) describes effect of the
spin-polarized current on the magnetic lattice under high spin injection.

Let us consider a configuration with $\mathbf H=\{0,\,0,\,H\}$, $\mathbf
n=\{0,\,0,\,1\}$, $\mathbf H_d=-4\pi M\{\hat M_x,\,0,\,0\}$, $\hat\mathbf
M_1=\{0,\,0,\,1\}$; $x$ axis being directed along the current, $yz$ plane
being coincided with the layer plane. On the spherical coordinates
$(\theta,\,\phi)$ with the polar axis along z axis the vector equation~(\ref{1})
determining direction of the free layer magnetization vector $\hat\mathbf
M=\{\sin\theta\cos\phi,\,\sin\theta\sin\phi,\, \cos\theta\}$ takes the
form of the following set of equations:

\begin{equation}\label{3}
  \frac{d\theta}{dT}=\frac{\sin\theta}{1+\kappa^2}\left\{-\kappa A(\theta,\,\phi,\,t)+B(\phi)\right\},
\end{equation}

\begin{equation}\label{4}
  \frac{d\phi}{dT}=\frac{1}{1+\kappa^2}\left\{A(\theta,\,\phi,\,t)+\kappa B(\phi)\right\},
\end{equation}
where

\begin{equation}\label{5}
  A(\theta,\,\phi,\,t)=h+h_a\cos\theta+\cos\theta\cos^2\phi+h_a\frac{j(t)}{j_0},
\end{equation}

\begin{equation}\label{6}
  B(\phi)=\cos\phi\sin\phi,
\end{equation}

\begin{displaymath}
  h=\frac{H}{4\pi M},\quad h_a=\frac{H_a}{4\pi M},\quad T=4\pi\gamma Mt.
\end{displaymath}

A numerical solution of Eqs.~(\ref{3}),~(\ref{4}) was found using Simulink software in the
MATLAB system~\cite{Karris}.

\begin{figure}
\includegraphics{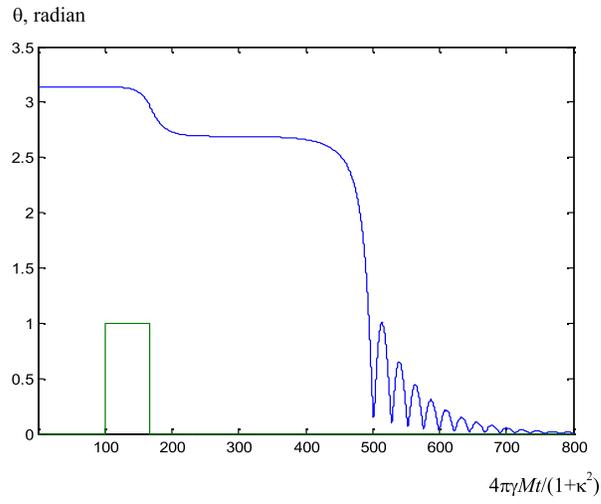}
\caption{Time dependence of the deviation angle of the free layer
magnetization vector from the initial antiparallel orientation under the
switching with joint action of the current pulse and the magnetic field.
$j/j_0=0.9$, $H/H_a=0.9$, $w=67.15t_0$.}\label{fig1}
\end{figure}

The following parameter values were given: $\kappa=0.03$, $h_a=0.01$. The
initial relative orientation of the pinned and free layers was assumed to
be antiparallel ($\theta=\pi$). At high enough current density and/or
magnetic field such a configuration becomes unstable, however, an initial
deviation is needed which is promoted by thermal fluctuations in reality.
The presence of the thermal noise creating deviation of the free layer
magnetization from the initial unstable stationary equilibrium state was
imitated with giving a small initial deviation from such a state by angle
of 0.01 radians in the layer plane where the demagnetization field does
not prevent fluctuation deviation, so that minimal fluctuation energy is
needed. The time dependence of the spin polarized current density $j(t)$
was given as a rectangular pulse of $w$ duration. The chosen values of the
magnetic field and the pulse amplitude, $H=0.9H_a$ and $j=0.9j_0$,
respectively, being $90\%$ of the corresponding threshold values, avoid
possibility of switching the junction by the field or by the current
singly. Therefore, the switching describing by the solution to be found is
caused by joint action of the current and magnetic field.

The simulation results are presented below as a time dependence of the
free layer deviation angle from the initial antiparallel orientation. The
dimensionless time is laid off as abscissa with
$t_0=(1+\kappa^2)/(4\pi\gamma M)$ as the time unit; at $M=900$ G one
nanosecond corresponds to 200 scale divisions of the abscissa ($t_0=0.005$
ns). The step shows the current turning on and turning off instants of time.

There is a minimal pulse duration under which the switching can occur yet
with the given values of the magnetic field and the pulse amplitude. In
Fig. 1 the free layer deviation angle from the initial antiparallel
orientation is shown as a function of time at $w=67.15t_0$. It is seen
that the switching takes place with substantial time delay relative to the
pulse. At the given values of the magnetic field and the current pulse
amplitude, that $w$ value is threshold one: with decreasing the pulse
duration only by $0.01t_0$ the switching does not occur (Fig. 2). Under
increasing in the pulse duration the switching delay time decreases (Fig.
3).

Lengthening the pulse makes it possible switching with lower values of the
magnetic field and/or the current density. In the limiting case of the
pulse duration much longer than the precession period, the switching condition
takes the form corresponding to the stationary current~\cite{Epshtein2},

\begin{equation}\label{7}
  \frac{H}{H_a}+\frac{j}{j_0}>1.
\end{equation}

\begin{figure}
\includegraphics{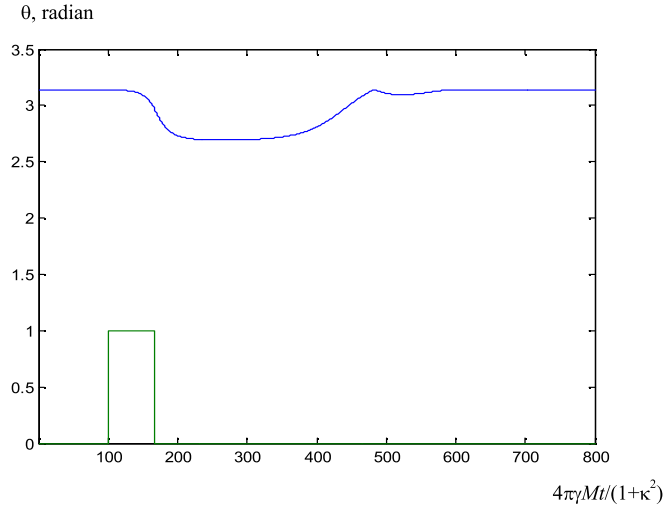}
\caption{Time dependence of the deviation angle of the free layer
magnetization vector from the initial antiparallel orientation with the
pulse duration below the threshold value. $j/j_0=0.9$, $H/H_a=0.9$,
$w=67.14t_0$.}\label{fig2}
\end{figure}

It is seen from Fig. 1 that the switching occurs as two stages, and the
longer stage is realized after the current turns off. This fact may be
explained by the following way. If the initial position of the
magnetization vector is not exactly parallel to the external magnetic
field, but is deviated from the antiparallel direction by some angle
$\psi$, then the magnetic field $H>H_a\cos\psi$ is needed for the
switching, the lower, the greater the deviation. The role of the current
pulse is that it, acting together with the magnetic field, deviates the
magnetization vector from the antiparallel direction by an angle large
enough that the magnetic field could do the junction switching alone.

\begin{figure}
\includegraphics{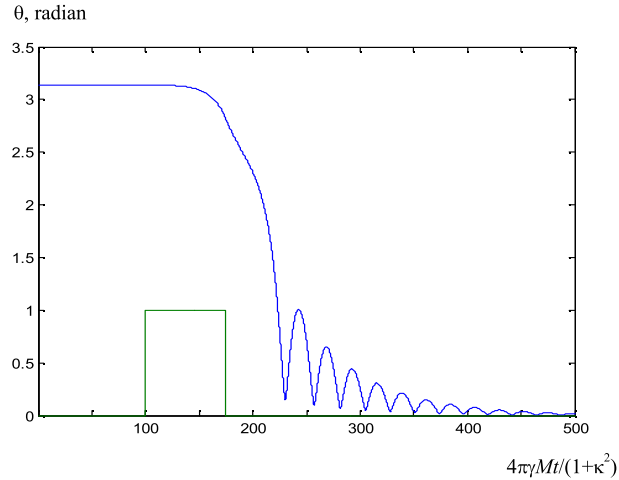}
\caption{Decreasing (cf. Fig.~\ref{fig1}) time delay of the switching under
pulse lengthening. $j/j_0=0.9$, $H/H_a=0.9$, $w=75t_0$.  }\label{fig3}
\end{figure}

The results show possibility of the switching magnetic junctions by joint
action of the magnetic field and current pulse. Varying the pulse
amplitude and duration allows obtaining, if necessary, a delayed switching
with controllable delay time.

The authors are grateful to Prof. Yu. G. Kusraev for useful discussions.

The work was supported by the Russian Foundation for Basic Research, Grant
No.~08-07-00290.


\begin{thebibliography}{12}
\bibitem{Katine}
J. A. Katine, F. J. Albert, R. A. Buhrman, E. B. Myers, D. C. Ralph, Phys. Rev. Lett.
\textbf{84}, 3149 (2000)
\bibitem{Diao}
Z. Diao, Z. Li, S. Wang, Y. Ding, A. Panchula, E. Chen, L.-C. Wang, Y.
Huai, J. Phys.: Condens. Matter \textbf{19}, 165209 (2007).
\bibitem{Watanabe}
M. Watanabe, J. Okabayashi, H. Toyao, T. Yamaguchi, J. Yoshino, Appl. Phys. Lett.
\textbf{92}, 082506 (2008).
\bibitem{Gulyaev1}
Yu. V. Gulyaev, P. E. Zil'berman, A. I. Krikunov, E. M. Epshtein, Zh.
Tekhn. Fiz. \textbf{77}, No. 9, 67 (2007) [Techn. Phys. \textbf{52}, 1169 (2007)].
\bibitem{Brataas}
A. Brataas, Y. V. Nazarov, G. E. W. Bauer, Eur. Phys. J. B \textbf{22}, 99 (2001).
\bibitem{Chung}
N. L. Chung, M. B. A. Jalil, S. G. Tan, J. Guo, S. Bala Kumar, J. Appl. Phys.
\textbf{104}, 084502 (2008).
\bibitem{Mankoff}
F. B. Mankoff, R. W. Dave, N. D. Rizzo, T. C. Eschrich, B. N. Engel, S. Tehrani,
Appl. Phys. Lett. \textbf{83}, 1596 (2003).
\bibitem{Chigarev}
S. G. Chigarev, E. M. Epshtein, P. E. Zilberman, Phys. Status Solidi B \textbf{247}, 325 (2010).
\bibitem{Gulyaev2}
Yu. V. Gulyaev, P. E. Zilberman, A. I. Panas, E. M. Epshtein, Zh.
Eksp. Teor. Fiz. \textbf{134}, 1200 (2008) [JETP \textbf{107}, 1027 (2008)].
\bibitem{Heide}
C. Heide, P. E. Zilberman, R. J. Elliott, Phys. Rev. B \textbf{63}, 064424 (2001).
\bibitem{Gulyaev3}
Yu. V. Gulyaev, P. E. Zilberman, E. M. Epshtein, R. J. Elliott,
Pis'ma Zh. Eksp. Teor. Fiz. \textbf{76}, 189 (2002) [JETP Letters \textbf{76}, 155 (2002)].
\bibitem{Epshtein1}
E. M. Epshtein, Yu. V. Gulyaev, P. E. Zilberman, J. Magn. Magn. Mater. \textbf{312}, 200 (2007).
\bibitem{Karris}
S. T. Karris, Introduction to Simulink with Engineering Applications (Orchard
Publications, 2006).
\bibitem{Epshtein2}
E. M. Epshtein, Radiotekh. Elektron. (Moscow) \textbf{54}, 339 (2009) [J. Commun. Technol. Electron.
\textbf{54}, 323 (2009)].
\end{thebibliography}
\end{document}